\documentclass[twocolumn,showpacs,preprintnumbers,amsmath,amssymb]{revtex4}
\usepackage{amsmath}
\usepackage{color}
\usepackage{graphicx}%Include figure files
\usepackage{dcolumn}%Align table columns on decimal point
\usepackage{bm}
\usepackage{overpic}

\begin{document}
\title{Effects of Levy Flights Mobility Pattern on Epidemic Spreading under Limited Energy Constraint}
\author{Yanqing Hu$^1$\footnote{yanqing.hu.sc@gmail.com}, Dan Luo$^1$, Xiaoke Xu$^2$, Zhangang Han$^1$, Zengru Di$^1$\footnote{zdi@bnu.edu.cn}}
 \affiliation{1. Department of Systems Science, School of Management and Center for Complexity
 Research, Beijing Normal University, Beijing 100875, China
\\2. School of Communication and Electronic Engineering, Qingdao Technological University, Qingdao, P. R. China}

\date{\today}%It is always \today, today,
             %but any date may be explicitly specified

\begin{abstract}
Recently, many empirical studies uncovered that animal foraging,
migration and human traveling obey Levy flights with an exponent
around -2. Inspired by the deluge of H1N1 this year, in this paper,
the effects of Levy flights' mobility pattern on epidemic spreading
is studied from a network perspective. We construct a spatial
weighted network which possesses Levy flight spatial property under
a restriction of total energy. The energy restriction is represented
by the limitation of total travel distance within a certain time
period of an individual. We find that the exponent -2 is the
epidemic threshold of SIS spreading dynamics. Moreover, at the
threshold the speed of epidemics spreading is highest. The results
are helpful for the understanding of the effect of mobility pattern
on epidemic spreading.

\end{abstract}


 %%%% check!!!!

\maketitle

%\section{Introduction}
%\label{1} With the globalization, the contact between people of
%different areas becomes more and more frequent. Consequently, most
%of epidemics are turning into challenges across regions and across
%countries. To study the characters and underlying mechanisms of
%epidemic spreading attracted many scientists' attentions. They can
%find applications to a wide range of problems, ranging from computer
%virus infections \cite{Email viru}, epidemiology such as the
%spreading of H1N1, SARS and HIV \cite{HIV}, and other spreading
%phenomena on communication and social networks, such as rumor
%propagation \cite{Rumor propagation}. Epidemic spreading always
%follows the mobility of  human and animal. Recently, many empirical
%studies and theoretic analysis presented that the mobility pattern
%possess Levy Flights' property that step size $d$ follows a
%power-law distribution $P(d)\sim d^{-\alpha} (1<\alpha\leq 3)$ with
%an exponent $\alpha \cong 2$ \cite{albatrosses, Random searches,
%predator,human travel,mobility}.
\section{Introduction}
With the process of globalization, the contacts between people of
different areas become much more frequent than before, which makes
the epidemic turn into an increasingly huge challenge for human
beings. In the recent 10 years, the worldwide deluges of epidemics
happened in human societies have been more frequent (e.g., the SARS
in 2003, the H5N1 in 2006 and H1N1 in 2009). The epidemic spreads
through interactions of human or even animals, and it appears more
powerful to damage as the interactions between people or animals get
stronger, under the current global environment, the effective
precaution actions should be made and taken according to available
studies. Complex networks, as models of interactions in many real
areas, such as society, technology and biology \cite{Review}, have
provided a practical perspective for studying the epidemic spreading
processes accompanying the real interactions. The spreading
processes can be regarded as dynamic processes on complex networks
\cite{Newman network,Review,Tao}. Hence, studying the
characteristics and underlying mechanisms of epidemic spreading on
complex networks, which can be applied to a wide range of areas,
ranging from computer virus infections \cite{Email viru},
epidemiology such as the spreading of H1N1, SARS and HIV \cite{HIV},
to other spreading phenomena on communication and social networks,
such as rumor propagation \cite{Rumor propagation}, has attracted
many scientists' attentions.

There are two main aspects of such studies. One is to aim at setting
the spreading mechanisms. For the epidemic spreading, the classical
models include SIS (susceptive-infected-susceptive) model and SIR
(susceptive-infected-remove) model\cite{Newman
network,Review,Tao,Vespignani}. The other one is focused on
researching the dynamic processes of epidemic spreading on networks
with different topological structures \cite{Newman
network,Review,Tao} which have an essential effect on the dynamic
processes of epidemic spreading. In regular, random and small-world
networks, the studies of epidemic spreading all found that the
dynamic processes undergo a phase transition: the effective
spreading rate needs to exceed a critical threshold for a disease to
become epidemic \cite{Newman network,Review,Small-world}. However,
accompanied by the discovery that more and more networks have a
scale-free distribution of degree, the absence of a critical
threshold has been revealed in the studies of epidemic spreading in
scale-free networks \cite{Scale-free PRL,endemic,heterogeneous}.

Epidemic spreading always follows the mobilities of human and
animal. Recently, many empirical studies and theoretical analysis on
the pattern of animal foraging, migration and human traveling have
presented that these mobility patterns possess a Levy flights
property with an exponent $\alpha\cong-2$ ($\alpha\cong-1.59$ for
human)\cite{predator, human travel, mobility}. Levy flights means
when human and animal travel, the step size $d$ follows a power-law
distribution $P(d)\sim d^{\alpha} (-3\leq\alpha<-1)$
\cite{Minireview}. Apparently, these mobility patterns cannot be
depicted only by the topology of networks; hence, the
characteristics of epidemic spreading following animal and human
mobility pattern should be observed on a network with specific
spatial structure. This work is to aim at abstracting a network to
describe the Levy flights mobility pattern and study the dynamic
process of epidemic spreading on spatial network.

In this paper we study how the mobility pattern affects epidemic
spreading from the network perspective, and especially, pay more
attention to the extremely rapid spreading epidemics. %There are two reasons why we use network approach. Firstly, network
%is easy to model and simulate. More importantly, the contacts
%between regions and countries are made up an immense complex
%network, therefore, it is reasonable to build a network model.
We construct a weighted network with Levy flights spatial structure
to describe the Levy flights mobility pattern. In this network, each
node denotes a small area and the weight on the edge denotes the
communication times or the quantities people or animal flow between
the corresponding two small areas. And, as all the individuals only
have limited energy, there must be a cost constraint on the
mobility. Considering this limitation, we let the consumed energy by
one communication between any two small areas be proportional to the
geographical distance between them.

%As the resource are limited, there is a
%restriction on total energy $\Omega$ in the model.

%After the mobility network established, we observed the diffusion of
%epidemics on it. To study the features of epidemic spreading, we
%considered the phase transition and the speed of spreading.

%Previous epidemic spreading studies based on the networks focused on
%the effects of the topological properties of networks and got a lot
%of conclusions\cite{Marc, Scale-free PRL, Phase in Lattice,
%Small-world, endemic, heterogeneous}. Topological property portrays
%the degree distribution of networks which can not reveal the
%networks' spatial structure sufficiently. As a result, our work is
%essentially different with those works and goes a significantly
%relevance to realistic perspective.

%\section{Mobility Network}\label{2}

\section{Mobility Network}
\begin{figure}
\center
\includegraphics[width=7cm]{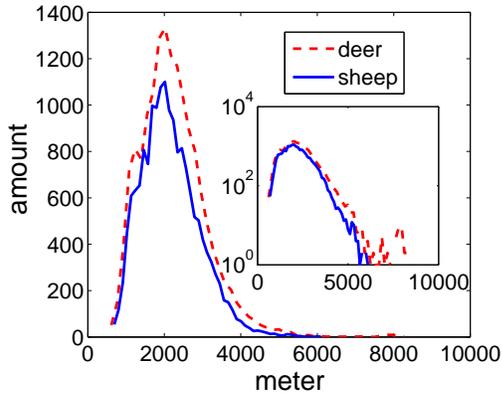}
\caption{The frequency distribution of sum of walk distances in a
day for deer and sheep. The data were obtained at the Macaulay
Institute's Glensaugh Research Station, Aberdeenshire, in north-east
Scotland. By equipping each one of 20 red deer and 14 sheep with a
GPS collar receiver instrument, which is fixed to the neck of the
animals, the data of locations during a total time period of 140
days were collected every 30 or 60 minutes. Both distributions are
very narrow and decay exponentially (subplots), which represents the
energy distribution is homogenous.}\label{empirical}
\end{figure}

We hypothesize that the distributions of human and animal mobility
energy are both homogenous. Then all nodes can be think than they
have the same energy, which can simplify the analysis considerably.
Unfortunately, we have no the data about human mobility and just
have deer and sheep moving data to support the hypothesis.  From
Fig. \ref{empirical}, we can see that the distributions of consumed
energy (the sum of distances for a sheep or deer in a day) for both
sheep and deer are very narrow, thus the energy distribution can be
looked as homogenous.

The spatial Levy flights network can be constructed as follows.
Based on a energy constraint uniform $n$ nodes cycle (1-dimensional
lattice), for each node, connections are added with power law
distance distribution $P(d)\sim d^{\alpha}$ randomly until the
energy are exhausted. Each realization can generate a specific
spatial weighted network. According to Levy flights mobility pattern
($P(d)\sim d^{\alpha})$, the weight $w_{ij}$ on the link between
node $i$ and $j$ should be proportional to $d_{ij}^{\alpha}$, and
for a given network size, the sum of all $w_{ij}d_{ij}$ should be a
constant which denotes the energy constraint. So, we can get an
ensemble network model of these spatial weighted networks generated
by many times of realization as:
\begin{eqnarray}
\label{aa}
\left\{
\begin{array}
{c}w_{ij}\sim d_{ij}^{\alpha} \\
\sum\limits_{i=1,j=1}^{n}w_{ij}d_{ij}=W
\end{array}\right.
\end{eqnarray}

Solve the model we can get the network as:
\begin{equation} w_{ij}=\frac{W d_{ij}^{\alpha}}{2nE(d)\sum_{d=1}^{\frac{n}{2}}d^{\alpha}}\end{equation}
Where, $W=wn$ denotes the total energy, $w$ is a constant and
$E(d)=\frac{\sum_{d=1}^{\frac{n}{2}}d^{1+\alpha}}{\sum_{d=1}^{\frac{n}{2}}d^{\alpha}}$
denotes the expectation of one Levy flights distance. Obviously, the
network is a full connected weighted network and each node is same
with the degree $k=\frac{W}{nE(d)}=\frac{w}{E(d)}.$

\section{Effects of Levy Flights Mobility on Epidemic Spreading}
So far we have constructed the one-dimensional Levy flights network.
SIS epidemiological network model is the standard model for studying
epidemic spreading \cite{Scale-free PRL}. In SIS endemic network,
each node has only two states, susceptive or infected. At each step,
the susceptive node $i$ is infected with rate $v$ if connected to
one infected node. Hence in the weighted network, the node $i$ will
be infected with the rate $1-(1-v)^{\sum_{j\in I}w_{ij}}$, where $I$
is the set of infected nodes. And at the next time, infected nodes
are cured and become susceptive with rate $\delta$. The effective
spreading rate $\lambda$, is defined as $\lambda=\frac{v}{\delta}$.
In order to keep the model simple and without loss of generality, we
always let $\delta=1$, which implies that all the infected nodes
will be cured in the next step. For many networks, the epidemic
threshold of $\lambda$ is a very significant index to measure the
dynamic processes of epidemic spreading on the networks \cite{Newman
network,Review,Scale-free PRL,endemic,heterogeneous}. Suppose
$\lambda_c$ is the epidemic threshold of a network which means that
when $\lambda<\lambda_c$ the infection dies out exponentially and
when $\lambda>\lambda_c$ the infection spreads and will always exist
in the network.

\begin{figure}
\center
\includegraphics[width=7.0cm]{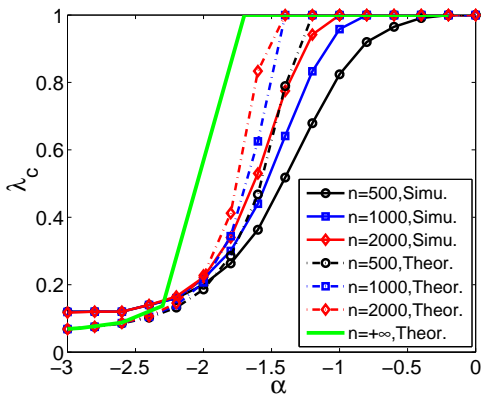}
\caption{The simulated and analytical dependence of $\lambda_c$ on
$\alpha$ under $w=10$. It shows the critical threshold $\lambda_{c}$
obtained by the numeric simulation model and reaction equations
under different exponent $\alpha$. Green line shows the analytical
critical threshold represented by Eq. (\ref{threshold}). As shown in
Eq. (\ref{threshold}), there is a phase transition on $\alpha=-2$.
When $\alpha\geq-2$, the epidemic threshold burst mounting up into
$1$, and when $\alpha<-2$, the epidemic threshold is small values.
Because when $n\rightarrow +\inf$, it is very hard to calculate
$\lambda_{c}$ around -2. We use a line to represent in the very
small region around -2.}\label{critical}
\end{figure}

Suppose $\rho$ denotes the fraction of infected node in the network,
following the homogeneous mean-field approximation \cite{Review},
the dynamical rate equations for the SIS model are
\begin{equation}\frac{d\rho}{dt}=-\rho+[1-(1-\lambda)^{\rho\frac{w}{E(d)}}](1-\rho)\label{rf}\end{equation}
The first term in Eq. (\ref{rf}), which is set $1$, is the recovery
rate of infected nodes. The second term takes into account the
expected probability that a susceptive node to get the infection on
this weighted network $[1-(1-\lambda)^{\rho\frac{w}{E(d)}}]$.
Employing Taylor expansions and ignoring the infinitesimal value of
higher order, when $\alpha<-2$ the epidemic threshold
$\lambda_{c}=\frac{1}{k}=\frac{E(d)}{w}.$ When $\alpha\geq-2$ and
$n\rightarrow+\infty$, $E(d)\rightarrow+\infty$. From the Eq.
(\ref{rf}), we have $\rho\frac{w}{E(d)}\rightarrow0$ for any $\rho$.
In order to keep $1-(1-\lambda)^{\rho\frac{w}{E(d)}}>0$, the
$\lambda_{c}$ must tend to $1$ when $n\rightarrow+\infty$. In
conclusion we have
\begin{eqnarray}
\label{threshold} \lambda_{c}=\left\{\begin{array}{c c} 1  &
\alpha\geq -2\\
\frac{1}{k}=\frac{E(d)}{w}\approx\frac{Z(\alpha)}{wZ(\alpha-1)} &
\alpha<-2,
\end{array}\right.
\label{th}\end{eqnarray} where
$Z(\alpha)=\sum_{d=1}^{+\infty}d^{\alpha}$ is Riemann Zeta function.
Eq. (\ref{th}) shows that $\lambda_c$ has a transition at
$\alpha=-2$. The simulated and analytical results are shown in Fig.
\ref{critical} which are match well. More over, the results can be
easily extended to high-dimensional space. The Eq. (\ref{threshold})
shows that the epidemic is liable to disappear ($\lambda_{c}=1$) if
there are more long-distance connections than short ones
($\alpha\geq-2$). This is an interesting and counterintuitive
conclusion. Intuitively, the epidemic will get more chances to
spread and consequently it will infect a wider range of people under
the condition that there are more long-distance connections.
However, under the restriction on the total energy, more
long-distance connections for a node mean a cost of a sharp decline
in the number of short-distance connections for exchange. Hence,
accompanied by the decrease of the interactions between nodes,
epidemic spreading becomes recession and liable to disappear.

For many epidemics such as HIV, if one is infected, he/she cannot be
cured and always has a possibility to infect others. While with many
other extremely rapid spreading epidemics such as H1N1 and SARS, we
always ignore them or have no antibiotics to control them at the
very beginning of spreading. In these conditions, SI model is a
reasonable model to study the spreading process. There is only one
difference between SIS and SI model. In SIS model, an infected node
will have a probability to become susceptive, while, in SI model,
when a node is infected, it cannot be cured and always will infect
other susceptive nodes. According to the outbreaks of H1N1 and SARS,
we know that there are always a few infected individuals and even
only one infected individual in some cases. So, in the following
numerical experiments, we preset only one node to be infected at the
beginning and the effective spreading rate $\lambda=0.05$ always.

\begin{figure}
\center
\includegraphics[width=4.5cm,height=3.9cm]{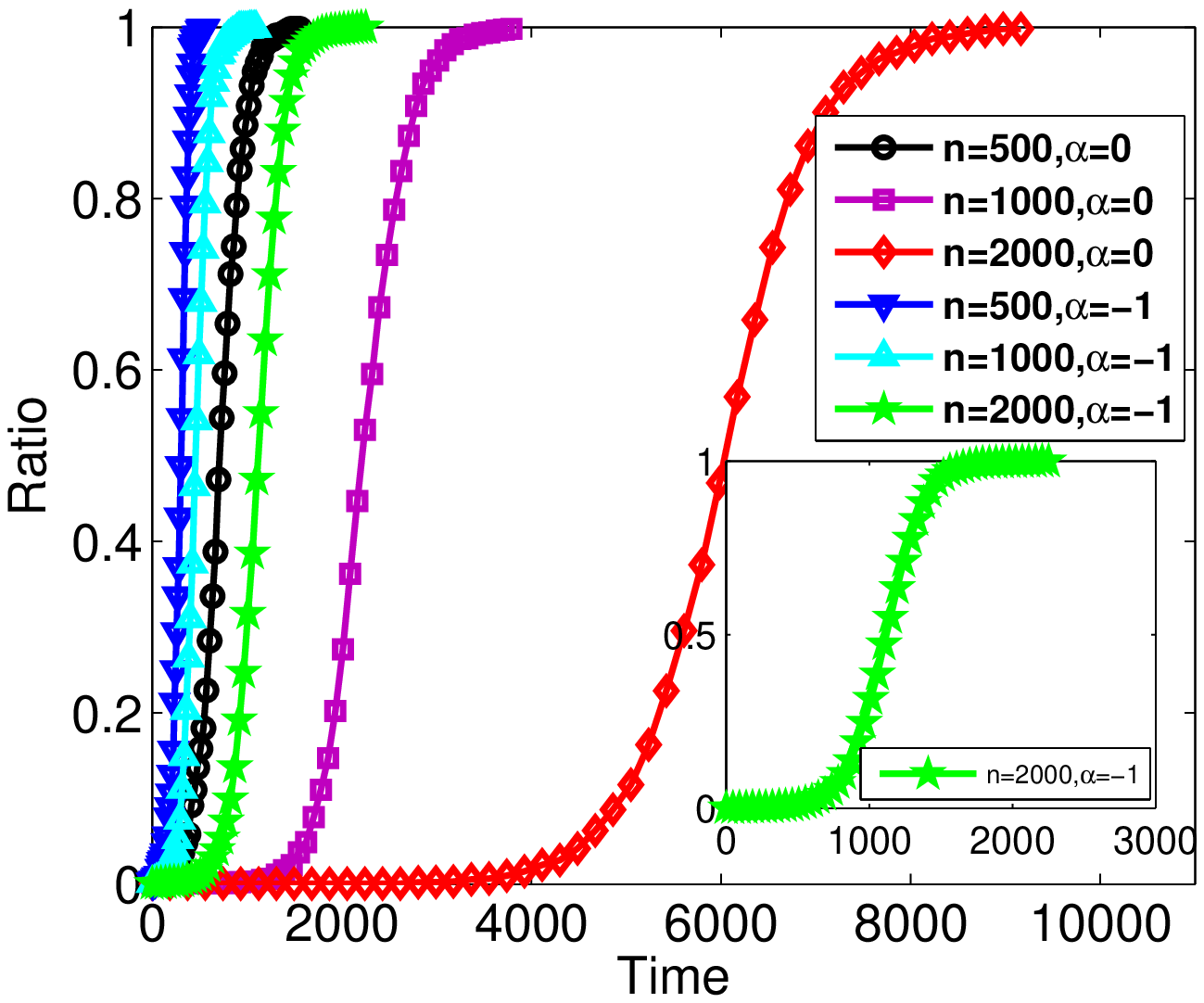}\includegraphics[width=4.5cm,height=3.9cm]{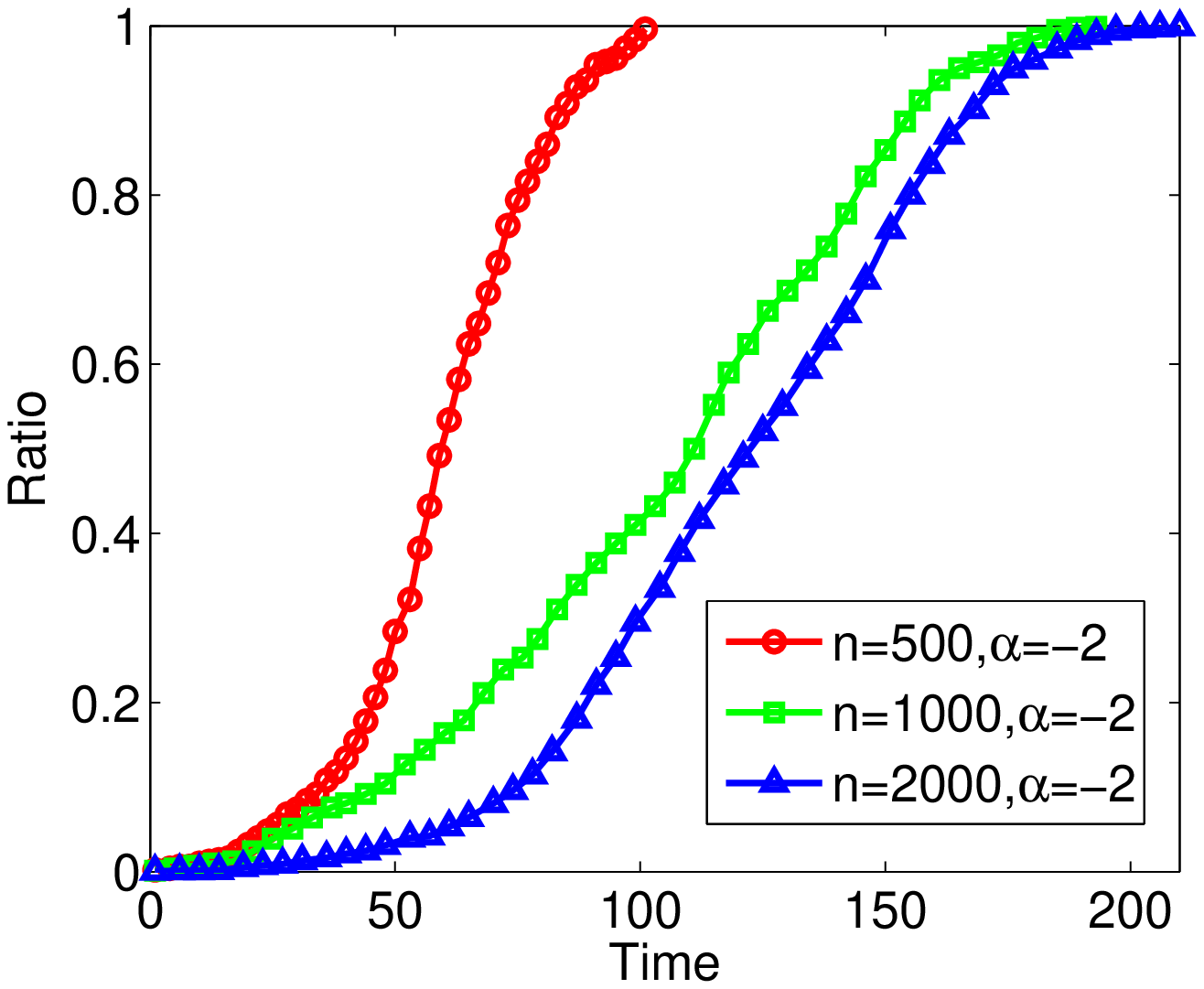}
\vskip -11.0em \hskip -26.0em  \raisebox{1em}{ \bf{A}} \vskip -2.0em
\hskip 1.8em { \bf{B}}\vskip 9.0em \hskip -20.0em
 \caption{The dependence of spread ratio on time under different $\alpha$. \textbf{A.} Spread ratio under
$\alpha=0$ and $\alpha=-1$, respectively. \textbf{B.} Spread ration
under $\alpha=-2$. We can see that when $\alpha=-2$, the spread
ratio grows much faster than other cases.} \label{spread ratio}
\end{figure}

\begin{figure}
\center
\includegraphics[width=7cm]{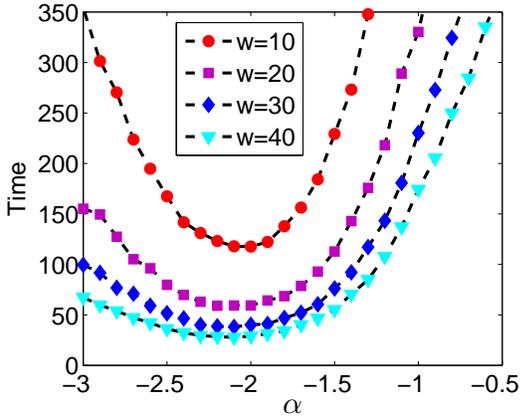}
\caption{The dependence of terminal time $T$ on $\alpha$ under
$n=1000$. It shows when the energy $w$ is not too large, the
terminal time $T$ achieves the lowest value around the point of
$\alpha=-2$, while the optimal $\alpha$ will deviate $-2$ and become
smaller when $w$ is very large.} \label{optimal}
\end{figure}

\begin{figure}
\center
\includegraphics[width=7cm]{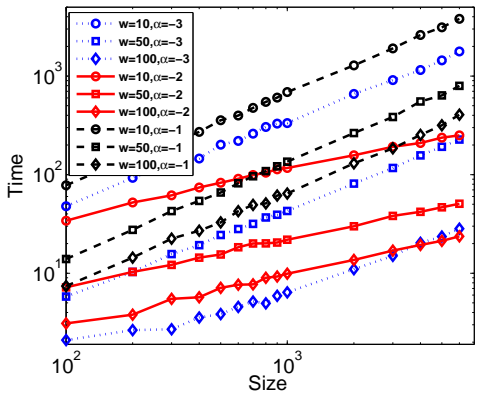}
\caption{The dependence of terminal time $T$ on network size $n$. It
observes the impact of network size on terminal time under various
exponent $\alpha$ and energy $w$. There is a log-log linear
dependence between terminate time and the network size. Move over we
find that, for not too large energy $w$, $T\propto n^{\beta}$. When
$\alpha=-2, \beta\approx\frac{1}{2}$.} \label{speed}
\end{figure}

The spreading processes are simulated on networks with different
network size. At first, we investigate the dependence of infected
ratio on spreading time. As shown in Fig. \ref{spread ratio}, we can
see that when $\alpha=-2$, the epidemic spread fastest under each
network size. Then, we also studied the dependence of terminal time
$T$ (at $T$, all nodes are infected) on $\alpha$ and we find that
when $w$ is not too large, $T$ achieves the lowest value around the
point of $\alpha=-2$ (shown in Fig. \ref{optimal}). Moreover for not
too large energy $w$, $T\propto n^{\beta}$. When $\alpha=-2$,
$\beta\approx\frac{1}{2}$ (shown in Fig. \ref{speed}).

\begin{figure}
\center
\includegraphics[width=7cm]{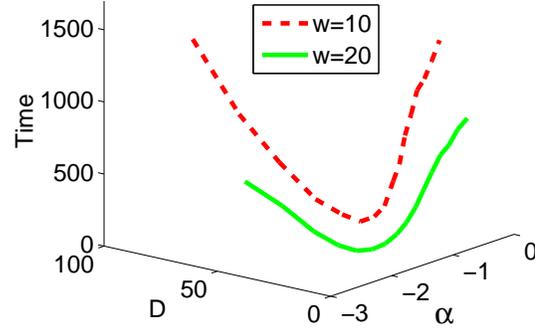}
\caption{The relationship among the terminate time $T$, the average
shortest path $D$ and Levy flights exponent $\alpha$. The lowest
point of the curve is close at $\alpha=-2$, at this point the
average shortest path and the terminate time are both the minimum
value.} \label{hh}
\end{figure}

\begin{figure}
\center
\includegraphics[width=7.2cm,height=6.2cm]{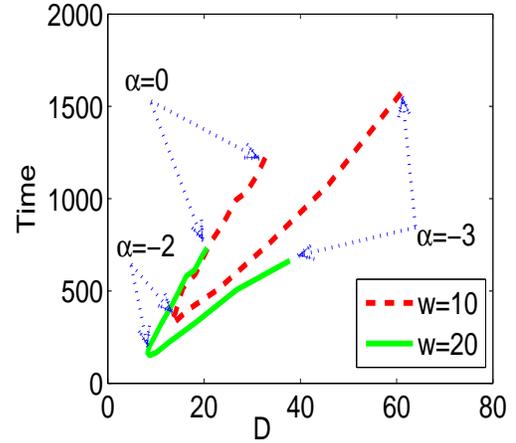}
\caption{The dependence of terminate time $T$ on average shortest
path $D$ under $n=2000$ with energy $w=10$ and $w=20$, the curve
shown the $T$ reach the minimum value when $D$ touch the lowest
point. Especially, when $\alpha>-2$, there is a linear relationship
between $D$ and $T$.} \label{hhh}
\end{figure}

In the spatial weighted network, the spreading process is the most
efficient when $\alpha=-2$. Can we get a similar conclusion in the
cases of higher dimensional? It is still an open question because it
seems tough to explain why epidemic spreading is fastest when
$\alpha=-2$ by using strict mathematical analysis, and the numeric
simulations on high-dimensional networks are too expensive to do.
However, we speculate this conclusion can be extended to the cases
of higher dimensional. Recent studies have presented that, with the
constraint $W=wn$, the average shortest path length $D$ also
achieves the lowest value when $\alpha=-2$ \cite{Hu,Halvin}. So, we
investigate the relationship between terminal time $T$ and the
average shortest path length $D$. However, it is not convenient to
obtain the average shortest path length for the spatial weighted
network. The reason is that large $w_{ij}$ means node $i$ and $j$
are very close, that is to say, we have to transform this kind of
weight if we want to obtain the shortest path which is of great
importance to us. In order to avoid transformation, we construct
another un-weighted network which is similar to spatial weighted
network. Based on a given uniform cycle network, we randomly add
connections which hold the above constraint and by avoiding
duplicated link to obtain an un-weighted network. By observing all
phenomena detected on the above spatial weighted network, we find
that the average shortest path length $D$ indeed affects the
spreading speed almost linearly, shown as in Fig. \ref{hh} and Fig.
\ref{hhh}, although, the slops at the two sides of $\alpha=-2$ are
slight different. This difference implies that there are some other
factors  which also affect the spreading speed besides the average
shortest path length. We believe that in high-dimensional networks
with Levy flights spatial structure, the epidemic also spreads
fastest when $\alpha=-2$. In the future we will try our best to
solve the question.

\section{Discussion and Conclusion}
In summary, this paper presents the relationship between epidemic
spreading and the mobility pattern of human and animal from the
network perspective. Acknowledging that the mobility pattern of
human and animal follows the Levy flights pattern with an exponent
$\alpha\cong-2$, we find that the Levy flights mobility pattern is
very efficient for epidemic spreading when $\alpha=-2$. This result
presents a big challenge for epidemics control nowadays. On the
other hand, this result in a sense complies with the evolution
theory. The special mobility pattern has evolved since animal
appeared on the earth. Maybe it is very useful for searching foods,
diffusing good genes or many others which are very important for
species developing. Consequently, it provides an effective path to
spread epidemic. So it is of no surprise to obtain such a
conclusion.

\textbf{Acknowledgement.}We wish to thank Prof. Shlomo Havlin and
Prof. Tao Zhou for some useful discussions. The work is partially
supported by NSFC under Grant No. 70771011 and 60774085. Y.Hu was
supported by the BNU excellent Ph.D Project.


\begin{thebibliography}{99}

\bibitem{Review} S. Boccaletti, V. Latora, Y. Moreno, M. Chavez, and
D. U. Hwang.\textit{ Physics Reports}. 424: 175-308 (2006).

\bibitem{Newman network}M. E. J. Newman.\textit{ Phys. Rev. E}. 66: 016128 (2002).
 \bibitem{Tao} T. Zhou, Zhong-Qian Fu, Bing-Hong Wang. \textit{Progress in Natural
 Science.} 16(5): 452-457 (2006).

\bibitem{Email viru}M. E. J. Newman, S. Forrest, and J. Balthrop.\textit{ Phys. Rev. E}. 66: 035101 (2002).

\bibitem{HIV}P. M. A. Sloot, C. Boucher. \textit{et al}. \textit{ Int. J. Comput. Math}. 85: 1175-1187  (2008).

\bibitem{Rumor propagation}D. H. Zanette. \textit{ Phys. Rev. E}. 65:
041908  (2002).


\bibitem{Vespignani} D. Balcan \textit{et al}. \textit{BMC Medicine}
7:45 (2009).

\bibitem{Small-world} M. Kuperman, G. Abramson. \textit{ Phys. Rev. Lett. } 86:2902  (2001).

\bibitem{Scale-free PRL}R. Pastor-Satorras, A. Vespignani. \textit{ Phys. Rev. Lett.} 86: 3200-3203  (2001).

\bibitem{endemic}R. Pastor-Satorras, A. Vespingnani. \textit{ Phys. Rev. E}. 63:
066117 (2002) .

\bibitem{heterogeneous}R. Pastor-Satorras, A. Vespingnani. \textit{The Europhys. Jour. B.} 26: 521-529 (2002).


\bibitem{predator}D. W. Sims, J. D. Metcalfe. \textit{et al}. \textit{Nature.} 451: 1098-1102 (2008).

\bibitem{human travel}D. Brockmann, L. Hufnagel, and T.
Geisel. \textit{Nature.} 439:462-465 (2006).

\bibitem{mobility}M. C. Gonzalez, C. A. Hidalgo, and
A.-L. Barabasi. \textit{Nature.} 453: 779-782 (2008).

\bibitem{Minireview}G. M. Viswanathan, H. E. Stanley. \textit{et al.} \textit{ Physica A.} 282: 1-12 (2000).

\bibitem{Hu} H. Yang, \textit{et al.} \textit{arXiv}: 0908.3968

\bibitem{Halvin}G. Li, \textit{et al.} (2009)
Designing optimal transport networks. \textit{arXiv}: 0908.3869.




\end{thebibliography}
\end{document}